\documentclass{article}
\usepackage{amsmath}
\usepackage{graphicx}

\begin{document}
\title{Renormalization of Hamiltonian QCD}
\author{A. Andra\v si\footnote{e-mail:aandrasi@rudjer.irb.hr} \\{\it 'Rudjer Bo\v skovi\' c' Institute, Zagreb,
  Croatia}\\ \\
\and  \\{John C. Taylor\footnote{e-mail:J.C.Taylor@damtp.cam.ac.uk}} \\{\it Department of Applied Mathematics and
  Theoretical Physics,}\\ \\{\it University of Cambridge, UK }\\}
\date{ April, 10 2007}
\maketitle

\section*{Abstract}
We study to one-loop order the renormalization of QCD in the Coulomb
gauge using the Hamitonian formalism. Divergences occur which might
require counter-terms outside the Hamiltonian formalism, but they
can be cancelled by a redefinition of the Yang-Mills electric field.
\vskip 1cm
PACS: 11.15.Bt; 11.10.Gh
\vskip 0.5cm
Keywords: Coulomb gauge, Hamiltonian, renormalization
\vfill \eject
\section{Introduction}
We study the renormalization of QCD in the Coulomb gauge Hamiltonian
formalism. By Hamiltonian form, we mean that the Lagrangian contains
only first order terms in time derivatives, and depends upon the conjugate
momentum field $E^a_i$ as well as the (transverse) gluon field $A^a_i$
(here $a$ is the colour index and $i=1,2,3$ is a 3-vector index).
This form has a number of attractive features:

(i) As a Hamiltonian exists, the theory is explicitly unitary, without
the necessity to cancel unphysical degrees of freedom with ghosts.

(ii) The Lagrangian form of the Coulomb gauge has ``energy divergences''
in some of its Feynman integrals, that is integrals of the form (we use $K$
for the spatial part of the 4-vector $k$)
$$\int d^3K dk_0 f(K,k_0) \eqno(1)$$
where $f$ does not decrease as $k_0 \rightarrow \infty$ (for fixed $K$).
These divergences cancel between different Feynman graphs \cite{Moh}, but this
cancellation has to be organized ``by hand''. In the Hamiltonian
form, each individual Feynman graph is free of such divergence.
Formally 'energy divergent' integrals such as
$$ \int{{d^3P}\over{(2\pi)^3}}\int
{{dp_0}\over{(2\pi)}}{{p_0}\over{p_0^2-P^2+i\eta}}\times
{1\over{(P-K)^2}} \eqno(2) $$
are assigned the value zero.

(iii) It has been argued \cite{Zwan} that the Coulomb gauge throws light
on confinement. Certainly it is known \cite{Fren} that, in the Coulomb
gauge, the source of asymptotic freedom lies in the Coulomb potential.

In spite of (i) above, to 2-loop order, mild energy-divergences
 remain \cite{Paul}, \cite{Doust}, \cite{Taylor} which result in ambiguities which have to be resolved
by a prescription. This is connected with questions of operator
ordering \cite{Christlee}. 

For other applications of the Coulomb
 gauge, for example to lattice QCD, see \cite{Cucczwan}, \cite{Cu}.

The question addressed here is the following.  Ultra-violet divergences exist
which seem to require the existence of counter-terms containing
second order terms in time derivatives, $(\partial A_i^a /\partial t)^2$.
Do these take us out of the Hamiltonian form? We argue that this
does not happen because the divergences concerned can be cancelled by a
redefinition of the $E_m^a$ field.

We do not use quite the strict Hamiltonian formalism. We retain the
auxilliary field $A_0^a$, which contains no time derivatives
and should be integrated out to give a nonlocal Coulomb
potential term in the real Hamiltonian. It seems to be
convenient, for the purposes of renormalization, to retain $A_0^a$
in the Lagrangian. Because of this, there is a  ghost field, but it has
an instantaneous propagator, and so is not relevant to unitarity.
Its purpose is only to cancel out closed loops in the $A_0^a$
field.

 \section{The Feynman rules}
The Lagrangian for the Coulomb gauge is
$$\mathcal{L'}=\mathcal{L}-{1\over 2\alpha}(\partial_i A_i^a)^2
\eqno(3)$$
(where $ \alpha $ will eventually tend to zero to go to the Coulomb
gauge),

$$\mathcal{L}=-{1\over 4} {\bf F}_{ij}\cdot
{\bf F}_{ij}-{1\over 2}
({\bf E}_i)^2+{\bf E}_i
\cdot {\bf F}_{0i}$$
$$+\partial _i{\bf c}^*\partial
 _i{\bf c}+g\partial
 _i{\bf c}^*\cdot({\bf A}_i\wedge
{\bf c})$$
 $$+{\bf u}_i\cdot
[\partial
  _i{\bf c}+g({\bf A}_i\wedge {\bf c})]$$
$$+{\bf u}_0\cdot[\partial
  _0{\bf c}+g({\bf A}_0\wedge{\bf c})]$$
$$-{1\over
 2}g{\bf K}\cdot({\bf c}\wedge
 {\bf c})+g{\bf v}_i\cdot({\bf E}_i\wedge
{\bf c})
 \eqno(4) $$
where we use a colour vector notation, and
$$  F^a_{ij}=\partial _i A^a_j-\partial _j A^a_i+gf^{abc} A^b_i A^c_j $$
and
$$ ({\bf A_i}\wedge
{\bf c})^a=f^{abd} A^b_i c^d \eqno(5) $$
Here ${\bf c}, {\bf c}^*$ are the ghost fields, and the sources ${\bf u}_i, {\bf v}_n$ and ${\bf K}$
are inserted for future use in formulating the BRST identities.
The conjugate momentum (electric) field ${\bf E}_m $ could be integrated out
to obtain the ordinary Lagrangian formalism, but for the Hamiltonian formalism
it must be retained.

We will use indices $m,n,...=1,2,3$ to denote the (spatial) components
of ${\bf E}$, so the seven fields are $( A^a_i,  A^a_0,  E^a_n)$. We will
use indices $I,J,..$ to denote the seven indices $(i,0,n)$.
The bilinear part of the Lagrangian in momentum space is a $7\times 7$
matrix
$$ S_{IJ}\delta_{ab}= \left( \begin{array}{rrr} $$-K^2(T_{ij}+L_{ij}/{\alpha})$$ & $$
  0$$ & $$ -ik_0\delta_{in} $$ \\  $$ 0 $$ &  $$ 0 $$ &  $$ iK_n $$ \\ $$
  ik_0\delta_{mj}$$ & $$ -iK_m $$ & $$ -\delta_{mn} $$ 
\end{array}   \right)  $$
where 
$$ T_{ij}\equiv \delta_{ij}-L_{ij}, {\hskip 1cm} L_{ij}\equiv
{K_iK_j}/{K^2}, $$
$$ k^2=k_0^2-K^2. \eqno(6) $$
For the propagators, we need the inverse
$$S^{-1}_{IJ}\delta_{ab}= \left( \begin{array}{rrr} $$T_{ij}/k^2 -
  \alpha L_{ij}/K^2 $$ & $$ \alpha k_0K_i/(K^2)^2 $$ & $$
  -ik_0T_{in}/k^2 $$ \\ $$ \alpha k_0K_j/(K^2)^2 $$ & $$ 1/K^2 +\alpha
  k_0^2/(K^2)^2 $$ & $$ iK_n/K^2 $$ \\ $$ ik_0T_{mj}/k^2 $$ & $$
  -iK_m/K^2 $$ & $$ T_{mn}K^2/k^2 $$ \end{array}  \right). \eqno(7)
  $$
We can now let $ \alpha \rightarrow 0 $, to obtain the Coulomb gauge.
From this, and the interaction terms in the Lagrangian (4), we can read off
the Feynman rules. We represent the ${\bf A}_i$ field by dashed lines,
the ${\bf E}_n$ field by continuous lines, and the ${\bf A}_0$ field by dotted lines.
With this notation, we now list the rules (a factor of $ {1\over {(2\pi)^4i}}$ is to be
included for each propagator, and a factor of $ (2\pi)^4i $ for each vertex). If we choose the
propagators in fig.1 to be the negative of the matrix (7), the extra factors of
$ {1\over{(2\pi)^4i}} $ for the propagator and $ (2\pi)^4i $ for the vertices cancel.

\section{The ultra-violet divergences}
The divergent graphs with 2 and with 3 external lines are shown in
Figures 4 till 31. Examples of the method of evaluation of divergent
parts are given in Appendices A and B.

The ultra-violet divergent parts of these graphs are, in terms of the
divergent constant (using dimensional regularization in $4-\epsilon$
dimensions) 
$$c= {{g^2}\over{16\pi^2}}C_G\Gamma(\epsilon/2), \eqno(8) $$
(where the superfix (4), (5) etc. refers to the corresponding figure
and $ \Pi_{ij},$ $ \Pi_{0i}$...$ \Pi_{mn} $ denote self-energies, $
V_{ijk},$ $ V_{0in}$...$ V_{0in} $ vertices and $ \Lambda $ stands for
diagrams with external ghost lines), are:
$$ \Pi^{(4)ab}_{ij}=ic[{1\over
    3}k_0^2\delta_{ij}+K^2\delta_{ij}-K_iK_j]\delta_{ab} \eqno(9) $$
$$ \Pi^{(5)ab}_{i0}=-{1\over 3}ick_0K_i\delta_{ab} \eqno(10) $$
$$ \Pi^{(6)ab}_{00}={1\over 3}icK^2\delta_{ab} \eqno(11) $$
$$ \Pi^{(7)ab}_{mi}=0 \eqno(12) $$
$$ \Pi^{(8)ab}_{m0}=-{4\over 3}ic[iK_i\delta_{ab}] \eqno(13) $$
$$ \Pi^{(9)ab}_{mn}=-{4\over 3}ic\delta_{mn}\delta_{ab} \eqno(14) $$
$$ V^{(10)abc}_{ijk}(p, q, r)=-{1\over
  3}cgf^{abc}[(Q-P)_k\delta_{ij}+(R-Q)_i\delta_{jk}+(P-R)_j\delta_{ik}]
  \eqno(15) $$
$$ V^{(11)abc}_{ijk}(p, q, r)=-{5\over
    6}cgf^{abc}[(Q-P)_k\delta_{ij}+(R-Q)_i\delta_{jk}+(P-R)_j\delta_{ik}]
    \eqno(16) $$
$$ V^{(12)abc}_{ijk}(p, q, r)=-{2\over
      3}cgf^{abc}[(Q-P)_k\delta_{ij}+(R-Q)_i\delta_{jk}+(P-R)_j\delta_{ik}]
      \eqno(17) $$
$$ V^{(13)abc}_{ijk}(p, q, r)={3\over
	2}cgf^{abc}[(Q-P)_k\delta_{ij}+(R-Q)_i\delta_{jk}+(P-R)_j\delta_{ik}]
	\eqno(18) $$
$$ V^{(14)abc}_{i00}(p, q, r)={1\over 4}cgf^{abc}(R-Q)_i \eqno(19) $$
$$ V^{(15)abc}_{i00}(p, q, r)=-{1\over 3}cgf^{abc}(R-Q)_i \eqno(20) $$
$$ V^{(16)abc}_{i00}(p, q, r)={1\over 3}cgf^{abc}(R-Q)_i \eqno(21) $$
$$V^{(17)abc}_{i00}(p, q, r)={1\over{12}}cgf^{abc}(R-Q)_i\eqno(22)$$
$$ V^{(18)abc}_{0jl}(p, q, r)=0 \eqno(23) $$
$$ V^{(19)abc}_{0jl}(p, q, r)={2\over 3}cgf^{abc}(R-Q)_0 \eqno(24) $$
$$ V^{(20)abc}_{0jl}(p, q, r)=-{1\over 3}cgf^{abc}(R-Q)_0 \eqno(25) $$
$$ V^{(21)abc}_{0jl}(p, q, r)=0. \eqno(26) $$
Graphs involving external $ {\bf E}_m $ line are
$$ V^{(29)abc}_{im0}(p, q, r)={1\over 3}icgf^{abc}\delta_{im}
\eqno(27) $$
$$ V^{(30)abc}_{im0}(p, q, r)=-{1\over 3}icgf^{abc}\delta_{im}
\eqno(28) $$
$$ V^{(31)abc}_{im0}(p, q, r)=0. \eqno(29) $$
All other graphs involving external $ {\bf E}_m $ -lines are convergent.
The divergent parts of graphs with open ghost line are
$$ \Lambda^{(22)ab}(q)=-{4\over 3}icQ^2\delta_{ab} \eqno(30) $$
$$ \Lambda^{(23)ab}_i(q)=-{4\over 3}cQ_i\delta_{ab} \eqno(31) $$
$$ \Lambda^{(24)abc}(p, q)=0 \eqno(32) $$
$$ \Lambda^{(25)abc}_k(p, q, r)=0 \eqno(33) $$
$$ \Lambda^{(26)abc}_0(p, q)=0 \eqno(34) $$
$$ \Lambda^{(27)abc}_i(p, q)=0 \eqno(35) $$
$$ \Lambda^{(28)abc}_n(p, q)=0. \eqno(36) $$

\section{Counter-terms}
Let
$$\Gamma_0 = \int d^4x\mathcal{L}(x) \eqno(37) $$
 be the original action, $\Gamma$ be the complete
effective action, and let $\Gamma_1$ be the effective action to
one-loop order. The complete BRST identities are
$$ \Gamma * \Gamma \equiv {{\partial \Gamma}\over{\partial
 {\bf A}_i}}\cdot {{\partial
 \Gamma}\over{\partial {\bf u}_i}}+{{\partial
 \Gamma}\over{\partial
 {\bf A}_0}}\cdot{{\partial
 \Gamma}\over{\partial {\bf u}_0}}+{{\partial
 \Gamma}\over{\partial
 {\bf c}}}\cdot{{\partial \Gamma}\over
{\partial {\bf K}}}+{{\partial
 \Gamma}\over{\partial
 {\bf E}_i}}\cdot{{\partial
 \Gamma}\over{\partial {\bf v}_i}}=0. \eqno(38)$$
So to one-loop order
$$\Gamma_1*\Gamma_0+ \Gamma_0 * \Gamma_1 \equiv \Delta \Gamma_1
=0 \eqno(39)$$
where
$$ \Delta={{\partial \Gamma}\over{\partial
    {\bf A}_i}}\cdot{{\partial}\over{\partial
    {\bf u}_i}}+{{\partial
    \Gamma}\over{\partial
    {\bf u}_i}}\cdot{{\partial}\over{\partial
    {\bf A}_i}}+{{\partial
    \Gamma}\over{\partial
    {\bf A}_0}}\cdot{{\partial}\over{\partial
    {\bf u}_0}}+{{\partial
    \Gamma}\over{\partial
    {\bf u}_0}}\cdot{{\partial}\over{\partial
    {\bf A}_0}} $$
$$+{{\partial \Gamma}\over{\partial
    {\bf c}}}\cdot{{\partial}\over{\partial
    {\bf K}}}+{{\partial \Gamma}\over{\partial
    {\bf K}}}\cdot
{{\partial} \over{\partial
    {\bf c}}}+{{\partial
    \Gamma}\over{\partial
    {\bf E}_i}}\cdot{{\partial}\over{\partial
    {\bf v}_i}}+{{\partial
    \Gamma}\over{\partial
    {\bf v}_i}}\cdot{{\partial}\over{\partial
    {\bf E}_i}} \eqno(40) $$
and
$$ \Delta^2=0. \eqno(41) $$

One class of solutions to this equation is of the form
$$ \Gamma_1^{(i)} = \Delta G, \eqno(42)$$
where the allowed form of $G$ is, in terms of constants $a_5,...a_{11}$,
$$
G=a_5{\bf A}_i\cdot({\bf u}_i+\partial _i
{\bf c^*})+a_6
	 {\bf A}_0\cdot
	 {\bf u}_0+a_7{\bf c}
\cdot
{\bf K}+a_8{\bf E}_i\cdot
 {\bf v}_i $$
$$+a_9{\bf v}_i\cdot \partial
 _i{\bf A}_0
+a_{10}{\bf v}_i\cdot \partial
 _0{\bf A}_i+a_{11}
{\bf v}_i\cdot({\bf A}_0\wedge
 {\bf A}_i). \eqno(43) $$
Other solutions of equation (39) are the explicitly gauge-invariant
terms
$$\Gamma_1^{(ii)} =
a_1({\bf F}_{ij})^2+a_2{\bf E}_i\cdot
{\bf F}_{0i}
+a_3({\bf F}_{0i})^2+a_4({\bf E}_i)^2.
\eqno(44) $$
Finally, by differentiating (38) with respect to the coupling constant $g$
and specialising to one-loop order, we see that
$$\Delta \Gamma_1^{(iii)}=0 \eqno(45)$$
where ($a_0$ being another divergent constant)
$$\Gamma_i^{(iii)} =a_0 g{\partial \Gamma_0 \over \partial
  g}.\eqno(46) $$

Combining these three contributions, we obtain
$$ \Gamma_1=\Gamma^{(i)}_1+\Gamma^{(ii)}_1+\Gamma^{(iii)}_1=\int d^4x
\mathcal{L}(x) \eqno(47) $$
where
$$
\mathcal{L}_1=a_1({\bf F}_{ij})^2+(a_2+a_8+a_9){\bf E}_i\cdot
{\bf F}_{0i} $$ $$ +(a_3-a_9)({\bf F}_{0i})^2
+(a_4-a_8)({\bf E}_i)^2 $$
$$+a_5{\bf F}_{ij}\cdot \partial
_j{\bf A}_i-(a_5+{1\over
  2}a_0)g{\bf F}_{ij}\cdot
({\bf A}_i\wedge
{\bf A}_j) $$
$$-(a_0+a_5+a_6)g{\bf E}_i\cdot({\bf A}_i\wedge 
{\bf A}_0)+{\bf E}_i\cdot(a_5\partial_0{\bf A}_i
-a_6\partial_i {\bf A}_0) $$
$$-a_5({\bf u}_i+\partial_i{\bf c}^*)\cdot \partial_i{\bf c} 
+a_0g\partial_i{\bf c}^*\cdot({\bf A}_i\wedge {\bf c})$$
$$-a_6{\bf u}_0\cdot \partial_0{\bf c}+a_0g{\bf u}_0 \cdot
({\bf A}_0\wedge
{\bf c}) $$
$$-a_7({\bf u}_i+\partial _i{\bf c}^*)\cdot
\{\partial_i{\bf c}+g({\bf A}_i\wedge{\bf c})\} $$
$$+a_0g{\bf u}_i\cdot({\bf A}_i\wedge
{\bf c})-a_7{\bf u}_0\cdot \{\partial_0{\bf c}
+g({\bf A}_0\wedge
{\bf c})\} $$ $$ +{1\over
  2}g(a_7-a_0){\bf K}\cdot
({\bf c}\wedge
{\bf c})+(a_0-a_7)g{\bf v}_i\cdot({\bf E}_i\wedge
 {\bf c}). \eqno(48) $$

The conditions coming from the vanishing ghost graphs Figs. 24, 25,
26, 27 and 28 are particularly simple. They fix
$$ a_9=-a_{10} $$
$$ a_{11}=-ga_9 $$
$$ a_0=a_7=-a_6.  \eqno(49) $$
In order for the counter-terms to cancel the divergences in the other
graphs, we require the conditions
$$ 4a_1-2a_5=-c $$
$$ 4a_1-3a_5-a_0={1\over 3}c $$
$$ a_3-a_9=-{1\over 6}c $$
$$ a_6-a_5={4\over 3}c $$
$$ a_5+a_7=-{4\over 3}c $$
$$ a_4-a_8={2\over 3}c $$
$$ a_2+a_5+a_8+a_9=0.  \eqno(50) $$

These equations do not fix the constants uniquely. We are free to make
some choices. The term $ ({\bf F}_{0i})^2 $ in $ \Gamma^{(ii)}_1 $ eq.(44)
is not present in the original Hamiltonian form of the Lagrangian (4),
so we choose
$$ a_3=0. \eqno(51) $$
We can also arrange for the combination
$$ -{1\over 2}({\bf E}_i)^2+{\bf E}_i\cdot {\bf F}_{0i}  \eqno(52) $$
to appear in $ \mathcal{L}^{(ii)}_1 $ as it does in $ \mathcal{L}_0. $
This requires (from (50))
$$ a_1=-{1\over 4}c+{1\over 2}a_5 $$
$$ a_2=c-2a_5 $$
$$ a_4=-{1\over 2}c+a_5 $$
$$ a_6={4\over 3}c+a_5 $$
$$ a_7=-{4\over 3}c-a_5 $$
$$ a_8=-{7\over 6}c+a_5 $$
$$ a_9={1\over 6}c $$
$$ a_0=-{4\over 3}c-a_5  \eqno(53) $$
and so
$$ \mathcal{L}^{(ii)}_1=-4a_1[-{1\over
    4}({\bf F}_{ij})^2-{1\over
    2}({\bf E}_i)^2+{\bf E}_i\cdot
    {\bf F}_{0i}]
\eqno(54) $$
proportional to the non-ghost part of the original Lagrangian (3).

Equation (54) does not come from the BRST identities, it just emerges
from the numerical values of the divergent integrals. It may be a
consequence of some hidden Lorentz invariance.

The constants $ a_0, a_1,...$ are still not uniquely fixed. There are
two particularly simple choices.

(i) Choose $ a_0=0 $ with $ a_5=-{4\over 3}c. $ Then we find
$$ a_1=-{{11}\over{12}}c $$
$$ a_2={{11}\over 3}c $$
$$ a_4=-{{11}\over 6}c $$
$$ a_6=a_7=0 $$
$$ a_8=-{5\over 2}c $$
$$ a_9={1\over 6}c.  \eqno(55) $$

(ii) The second choice is $ a_1=0 $ with $ a_5={1\over 2}c .$ Then
$$ a_0=-{{11}\over 6}c $$
$$ a_2=0 $$
$$ a_4=0 $$
$$ a_6={{11}\over 6}c $$
$$ a_7=-{{11}\over 6}c $$
$$ a_8=-{2\over 3}c $$
$$ a_9={1\over 6}c. \eqno(56) $$
Note that $ a_0 $ has the expected value for coupling constant
renormalization.

The counter-terms in either case are

$$
\mathcal{L}_1=-{{11}\over{12}}c({\bf F}_{ij})^2-{4\over
  3}c{\bf F}_{ij}\cdot \partial
_j{\bf A}_i+{4\over
  3}cg{\bf F}_{ij}\cdot ({\bf A}_i\wedge {\bf A}_j) $$
$$ -{1\over 6}c({\bf F}_{0i})^2+{2\over
    3}c({\bf E}_i)^2
+{4\over 3}c{\bf E}_i\cdot
    {\bf F}_{0i} $$
$$+{4\over
      3}cg{\bf E}_i\cdot ({\bf A}_i\wedge
      {\bf A}_0)
-{4\over 3}c{\bf E}_i\cdot
      \partial_0{\bf A}_i
+{4\over
      3}c({\bf u}_i+\partial_i{\bf c}^*)\cdot \partial_i
{\bf c}.  \eqno(57) $$

The counter-terms in $ a_5, a_6, a_7, a_8 $ and $ a_9 $ are involved
in a rescaling of the fields. Defining
$$ {\bf A}'_i=(1+a_5){\bf A}_i $$
$$ {\bf A}'_0=(1+a_6){\bf A}_0 $$
$$ {\bf E}'_m=(1+a_8){\bf E}_m-a_9{\bf F}_{0m} $$
$$ {\bf u}'_i=(1-a_5){\bf u}_i $$
$$ {\bf u}'_0=(1-a_6){\bf u}_0 $$
$$ {\bf c}'=(1-a_7){\bf c} $$
$$ {\bf K}'=(1+a_7){\bf K} $$
$$ g'=(1+a_0)g $$
$${\bf c}'^*=(1-a_5){\bf c}^* $$
$$ {\bf v}'=(1-a_8){\bf v},  \eqno(58) $$
we have from (48) that
$$ \mathcal{L}_0+\mathcal{L}_1=(1-4a_1)\mathcal{L}_0(g', {\bf A}'_i, {\bf A}'_0,
{\bf E}', {\bf c}', {\bf c}'^*, {\bf u}'_i, {\bf u}'_0, {\bf K}'). \eqno(59) $$

Note that $ a_6 $ which determines the renormalization of the Coulomb
field $ A^a_0 $ has the same numerical value as $ a_0 $. 

We have not calculated the divergences in graphs with four external lines. We
assume they will be cancelled by the same counter-terms.

\section {Comments}
We conclude that there is no difficulty to one-loop order in renormalizing the
Hamiltonian form of the Coulomb gauge. We guess that the renormalization
would formally go through to higher orders, but then there is the problem
mentioned in \cite{Paul}, \cite{Doust}, \cite{Taylor} of combining the renormalization of ultra-violet
divergences with the resolution of energy-divergence ambiguities.

It is not quite obvious how the renormalization would be formulated
if the $ A_0^a$ field had been eliminated to give the non-local
colour Coulomb potential (note the non-zero value of the $ A_0^a$
field renormalization constant $a_6$ in (56)). 
\vskip 0.5cm

Acknowledgements

A.A. wishes to thank the Royal Society for a grant and DAMTP for
hospitality.
We are grateful to Dr. G. Duplan\v ci\' c for drawing the figures.
The work was supported by the Ministry of Science and Technology of
the Republic of Croatia under contract No. 098-0000000-2865.
\vskip 0.5cm

\appendix {Appendix A}

Here we give as an example the evaluation of the ultra-violet
divergent part of the graph in Fig. 20.
$$ V^{(20)abc}_{0jk}(q, -q, 0)=ig^3C_Gf^{abc}\int
d^4p{{p_0P_k}\over{(p^2+i\eta)^2}}\cdot{1\over{(q+p)^2+i\eta}}$$
$$ \times T_{rz}(P)T_{zv}(P)T_{ru}(Q+P)
[(-2Q-P)_v\delta_{uj}+(Q-P)_u\delta_{jv}+(2P+Q)_j\delta_{vu}].
\eqno(A1) $$
Applying the integral
$$\int dp_0{{p_0}\over{(p^2+i\eta)^2}}{1\over{(q+p)^2+i\eta}} $$
$$=i\pi^{{1\over 2}}\Gamma({5\over 2})q_0\int_{0}^{1}dy
y(1-y)\{(P+yQ)^2+y(1-y)(-q^2-i\eta)\}^{-{5\over 2}} \eqno(A2) $$
and power counting to (A1)
$$ V^{(20)abc}_{0jk}(q, -q, 0)=-4g^3C_Gf^{abc}\sqrt\pi
q_0\Gamma({5\over 2})\int_{0}^{1}dy y(1-y) $$  $$ \int
d^{3-\epsilon}PP_jP_k\{(P+yQ)^2+y(1-y)(-q^2-i\eta)\}^{-{5\over 2}},
  \eqno(A3) $$ 
leading to
$$ V^{(20)abc}_{0jk}(q, -q, 0)=-{1\over
  3}cgf^{abc}q_0\delta_{jk}. \eqno(A4) $$
\vskip 0.5cm

\appendix {Appendix B}

Example of self-energy evaluation $ \Pi^{(6)ab}_{00} $ in eq.(11).
Let $ p, q $ be internal and $ k $ external momentum, $ p-q=k. $ The sum of two graphs is
$$ (2\pi)^{-4}{{T_{ij}(P)T_{ji}(Q)}\over{p^2q^2}}[{1\over
    2}(P^2+Q^2)-(ip_0)(iq_0)]\delta_{ab} \eqno(B1) $$
where we have symmetrized the first term in $ P, Q. $
the minus sign in the second term comes from the opposite order of the
 $ f^{abc} $ factors at the two vertices. Doing the $ p_0 $
    integration by Cauchy, we get
$$ (2\pi)^{-4}(2\pi
    i){{T_{ij}T_{ji}}\over{4PQ}}{1\over{(P+Q)^2-k_0^2}}(P+Q)[P^2+Q^2-2PQ]\delta_{ab}.
    \eqno(B2) $$
The last factor $ (P-Q)^2 $ is approximately $ (P\cdot K)^2/P^2. $
With this factor, the integral is only logarithmically divergent, and
    to get the divergent part we can put $ Q=P $ everywhere. We use $
    T_{ij}(P)T_{ji}(P)=2. $ Then we get
$$ (2\pi)^{-4}{{2\pi i}\over 4}K_iK_j \int
    d^{3-\epsilon}P{{P_iP_j}\over{(P^2+m^2)^{5/2}}}. \eqno(B3) $$
So the divergent part is\footnote{Note that there was an error of sign
    in Eur. Phys. J. C37, 307-313(2004) which however did not
    influence the final result.}
$$ {1\over 3}icK^2\delta_{ab}. \eqno(B4) $$

\bibliographystyle{unsrt}

\begin{figure}
\begin{center}
\includegraphics[width=8cm]{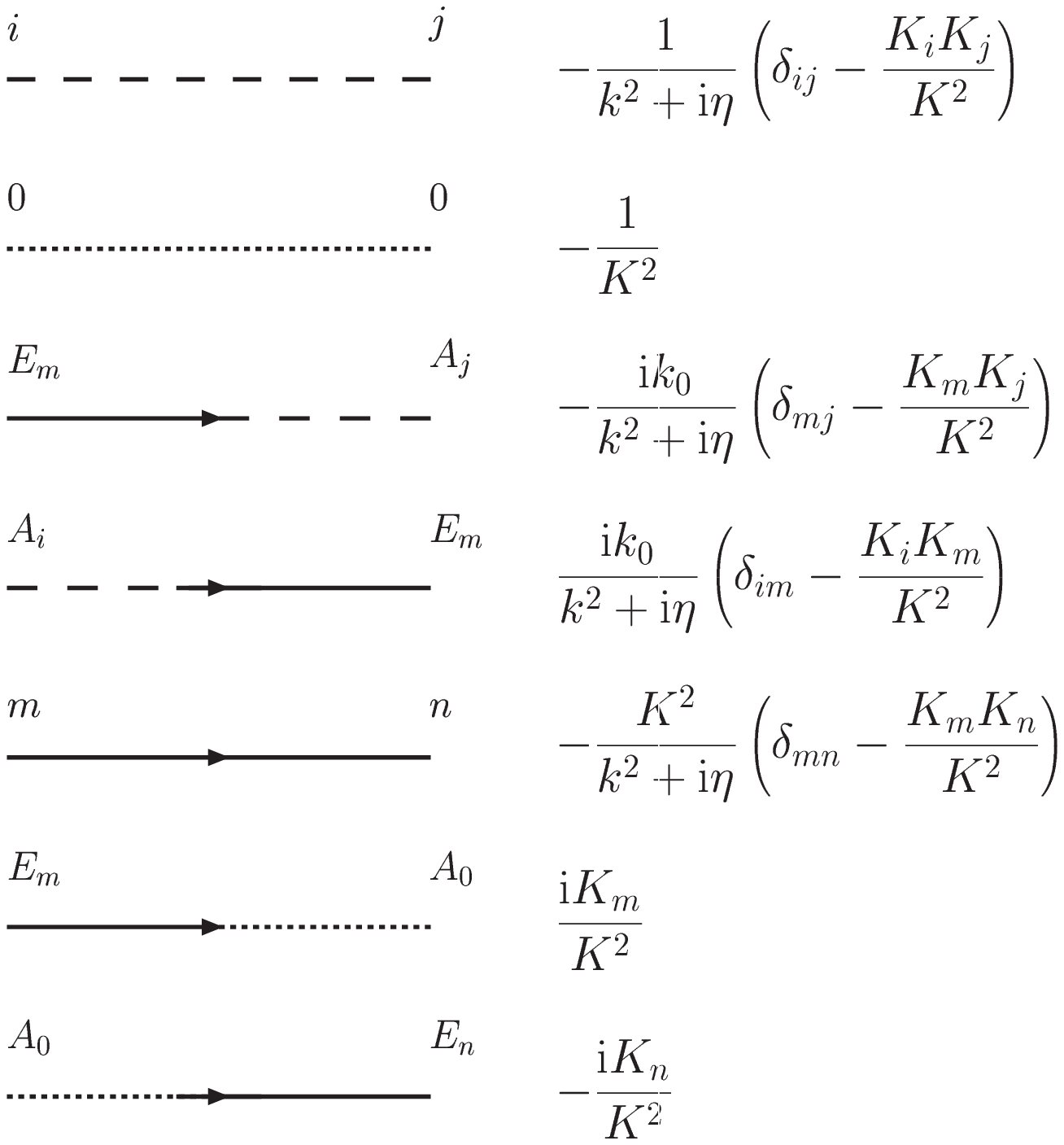}%
\caption{Feynman rules for the propagators in the Coulomb gauge.}
\end{center}
\end{figure}

\begin{figure}
\begin{center}
\includegraphics[width=8cm]{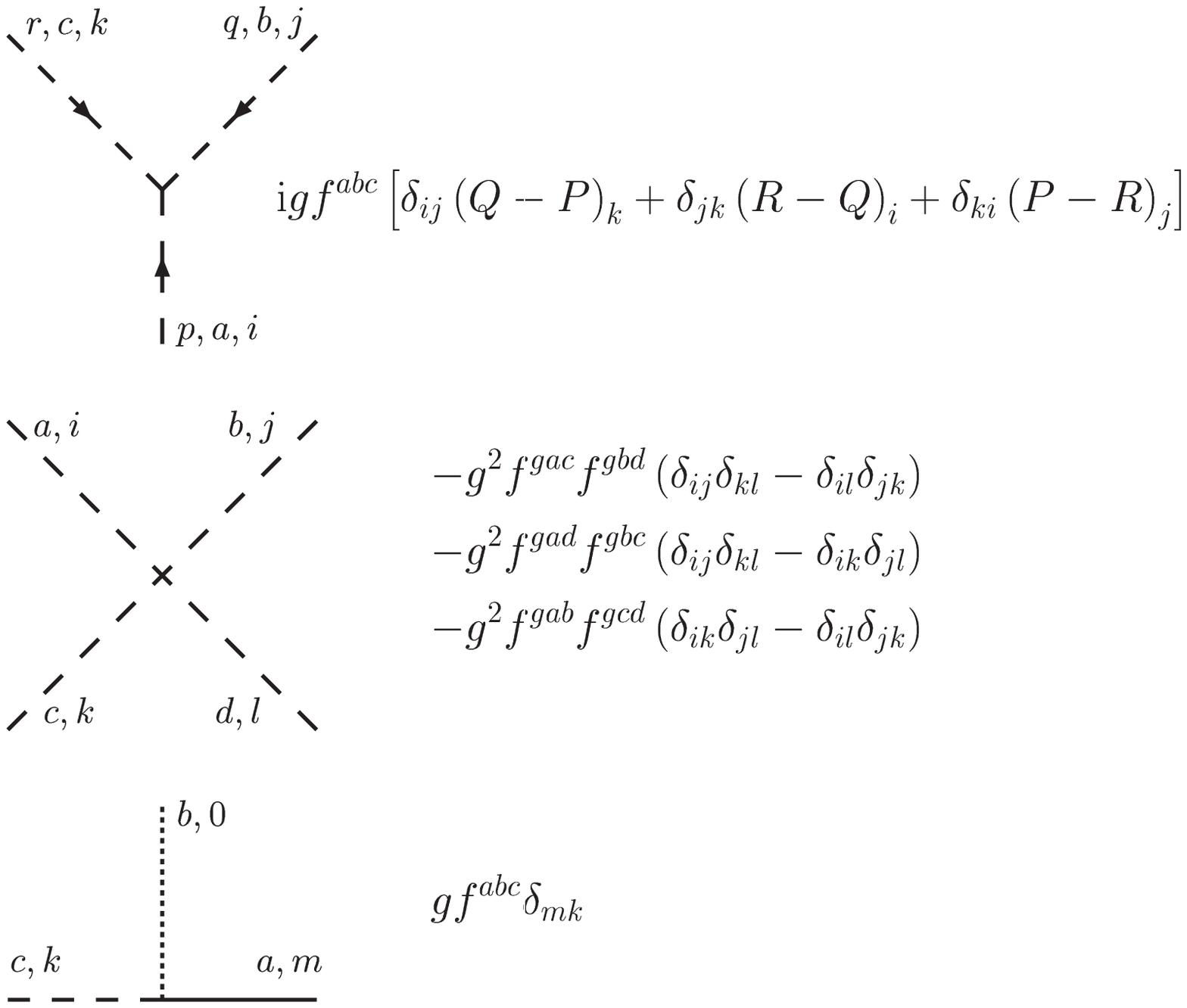}%
\caption{Feynman rules for the vertices in the Coulomb gauge. The arrows
denote the directions of the momenta.}
\end{center}
\end{figure}

\begin{figure}
\begin{center}
\includegraphics[width=6.2cm]{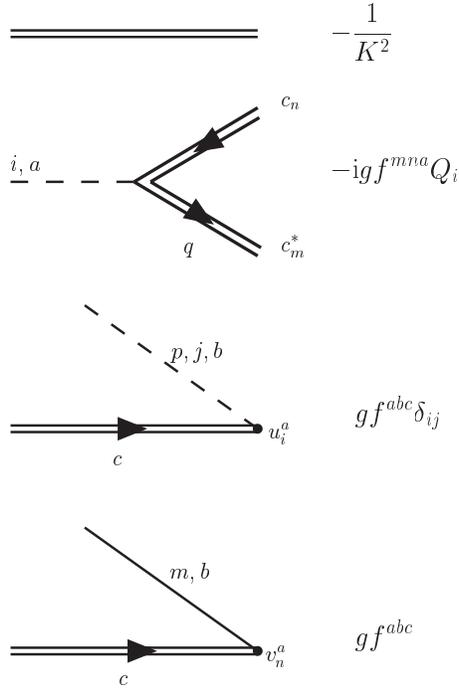}%
\caption{Feynman rules for ghosts and sources in the Coulomb gauge. Doubled 
lines denote ghosts. The black arrows distinguish between ghosts and antighosts.
Momenta flow into the vertex.}
\end{center}
\end{figure}

\begin{figure}
\begin{center}
\includegraphics[width=7cm]{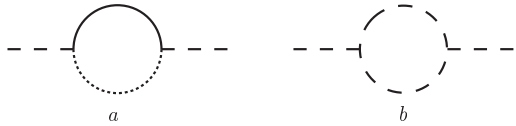}%
\caption{The transverse gluon self-energy graphs.}
\end{center}
\end{figure}

\begin{figure}
\begin{center}
\includegraphics[width=7cm]{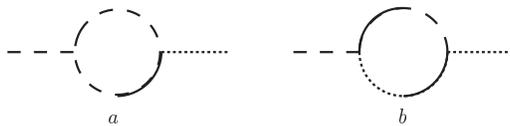}%
\caption{The $ {\bf A}_i{\bf A}_0 $ two-point function.}
\end{center}
\end{figure}

\begin{figure}
\begin{center}
\includegraphics[width=7cm]{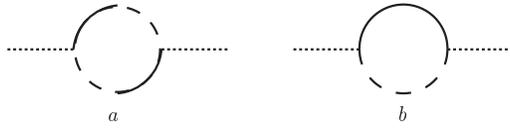}%
\caption{The time-time component of the gluon self-energy.}
\end{center}
\end{figure}

\begin{figure}
\begin{center}
\includegraphics[width=5.5cm]{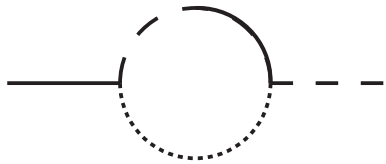}%
\caption{The transition between the transverse gluon field and its conjugate
 field $ {\bf E}_i $.}
\end{center}
\end{figure}

\begin{figure}
\begin{center}
\includegraphics[width=5.5cm]{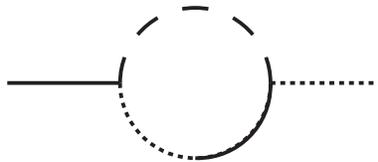}%
\caption{The transition between the Coulomb field $ {\bf A}_0 $ and 
the conjugate field $ {\bf E}_i $. }
\end{center}
\end{figure}

\begin{figure}
\begin{center}
\includegraphics[width=5.5cm]{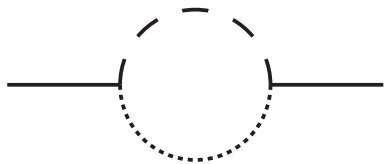}%
\caption{The conjugate field self-energy.}
\end{center}
\end{figure}

\begin{figure}
\begin{center}
\includegraphics[width=5.5cm]{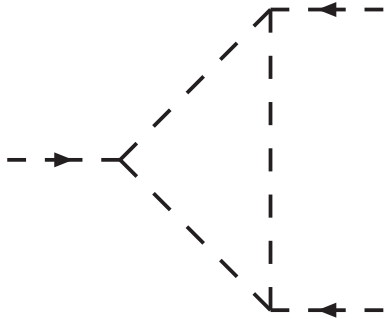}%
\caption{Graph contributing to the three-gluon vertex function.}
\end{center}
\end{figure}

\begin{figure}
\begin{center}
\includegraphics[width=5.5cm]{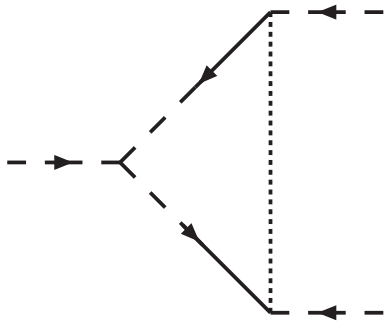}%
\caption{There are three graphs in this class with permutations of the vertices.}
\end{center}
\end{figure}

\begin{figure}
\begin{center}
\includegraphics[width=5.5cm]{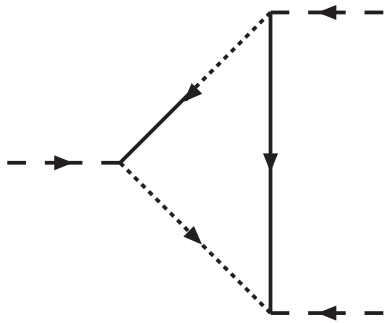}%
\caption{Graph representing a class of 6 diagrams.}
\end{center}
\end{figure}

\clearpage

\begin{figure}
\begin{center}
\includegraphics[width=5.5cm]{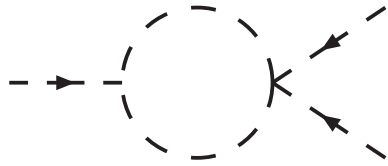}%
\caption{There are 3 graphs in this class of diagrams.}
\end{center}
\end{figure}

\begin{figure}
\begin{center}
\includegraphics[width=5.5cm]{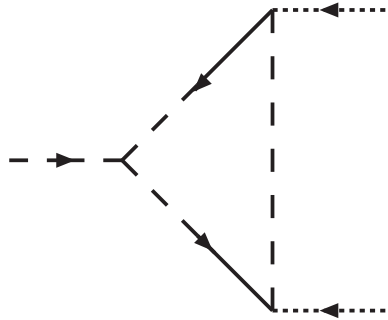}%
\caption{ Graph with two external Coulomb lines (there are 3 diagrams in this class).}
\end{center}
\end{figure}

\begin{figure}
\begin{center}
\includegraphics[width=5.5cm]{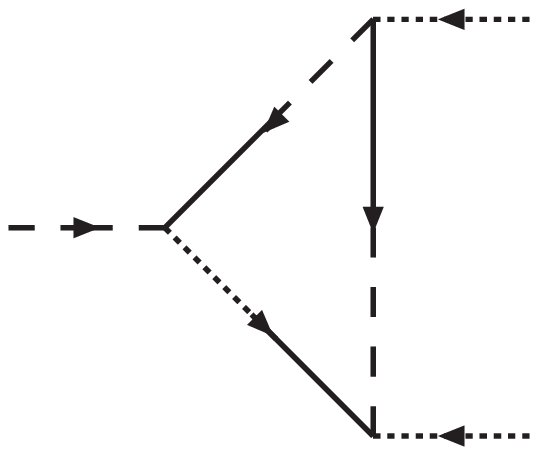}%
\caption{ There are two graphs in this class.}
\end{center}
\end{figure}

\begin{figure}
\begin{center}
\includegraphics[width=5.5cm]{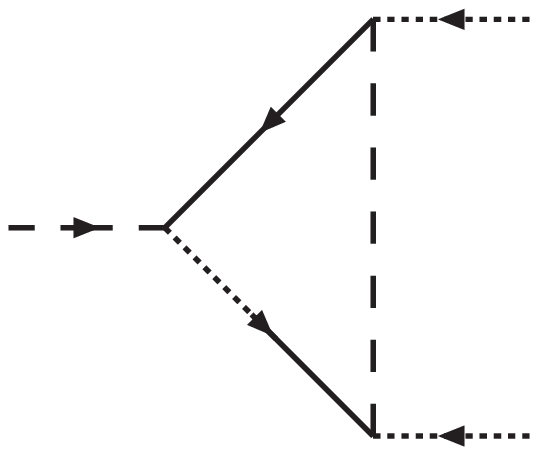}%
\caption{There are two graphs in this class.}
\end{center}
\end{figure}

\begin{figure}
\begin{center}
\includegraphics[width=5.5cm]{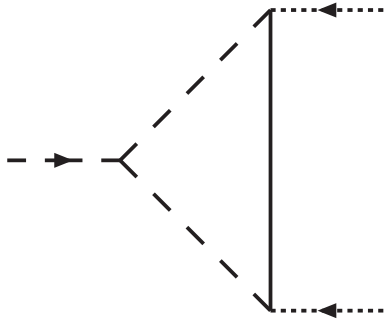}%
\caption{The graph with two external Coulomb lines and one three-gluon vertex.}
\end{center}
\end{figure}

\begin{figure}
\begin{center}
\includegraphics[width=8cm]{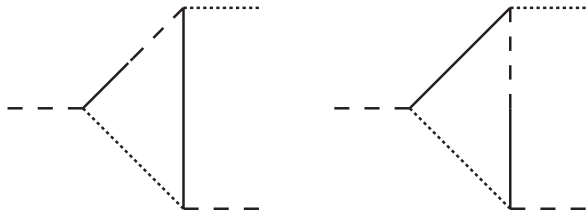}%
\caption{Graphs contributing to the$ ({\bf A}_i{\bf A}_j{\bf A}_0) $ three-point function.}
\end{center}
\end{figure}

\begin{figure}
\begin{center}
\includegraphics[width=5.5cm]{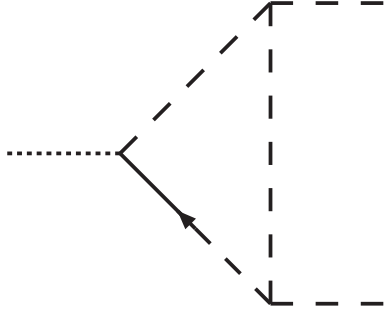}%
\caption{Graph contributing to the $ ({\bf A}_i{\bf A}_j{\bf A}_0) $ three-point
  function which contains a three-gluon vertex.}
\end{center}
\end{figure}

\begin{figure}
\begin{center}
\includegraphics[width=5.5cm]{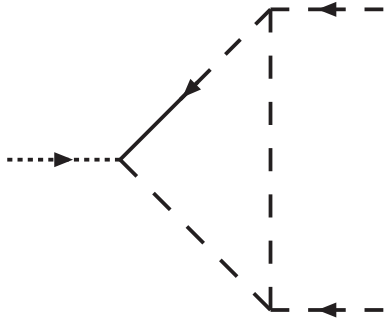}%
\caption{The $ ({\bf A}_i{\bf A}_j{\bf A}_0 ) $ graph with a three-gluon vertex.}
\end{center}
\end{figure}

\begin{figure}
\begin{center}
\includegraphics[width=5.5cm]{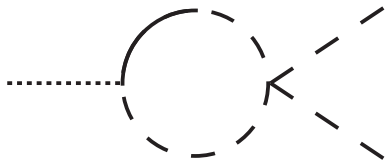}%
\caption{The $ ({\bf A}_i{\bf A}_j{\bf A}_0) $ graph with a four-gluon vertex.}
\end{center}
\end{figure}

\begin{figure}
\begin{center}
\includegraphics[width=5.5cm]{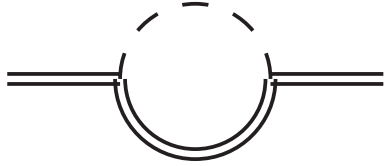}%
\caption{The ghost self-energy.}
\end{center}
\end{figure}

\begin{figure}
\begin{center}
\includegraphics[width=5.5cm]{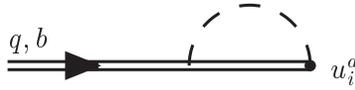}%
\caption{Ghost and the $ {\bf u}_i $ source graph.}
\end{center}
\end{figure}

\begin{figure}
\begin{center}
\includegraphics[width=5.5cm]{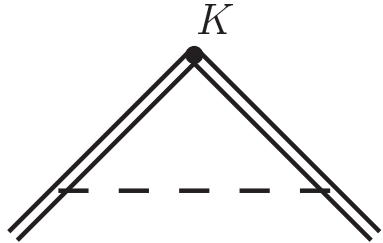}%
\caption{The ghost vertex graph with a $ {\bf K} $ source.}
\end{center}
\end{figure}

\clearpage

\begin{figure}
\begin{center}
\includegraphics[width=5.5cm]{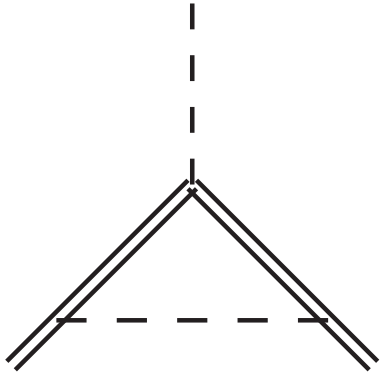}%
\caption{Graph with external $ {\bf A}_i, $ ghost and anti-ghost lines.}
\end{center}
\end{figure}

\begin{figure}
\begin{center}
\includegraphics[width=5.5cm]{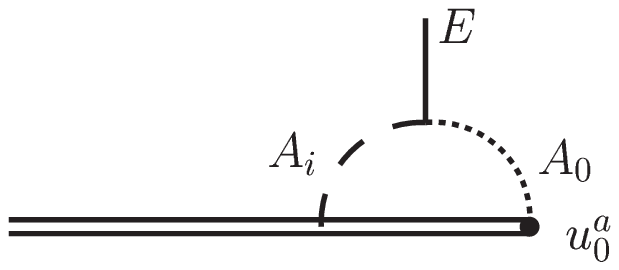}%
\caption{Graph with $ {\bf u}_0 $ source, $ {\bf E}_i $ and $ {\bf c} $ lines.}
\end{center}
\end{figure}

\begin{figure}
\begin{center}
\includegraphics[width=8cm]{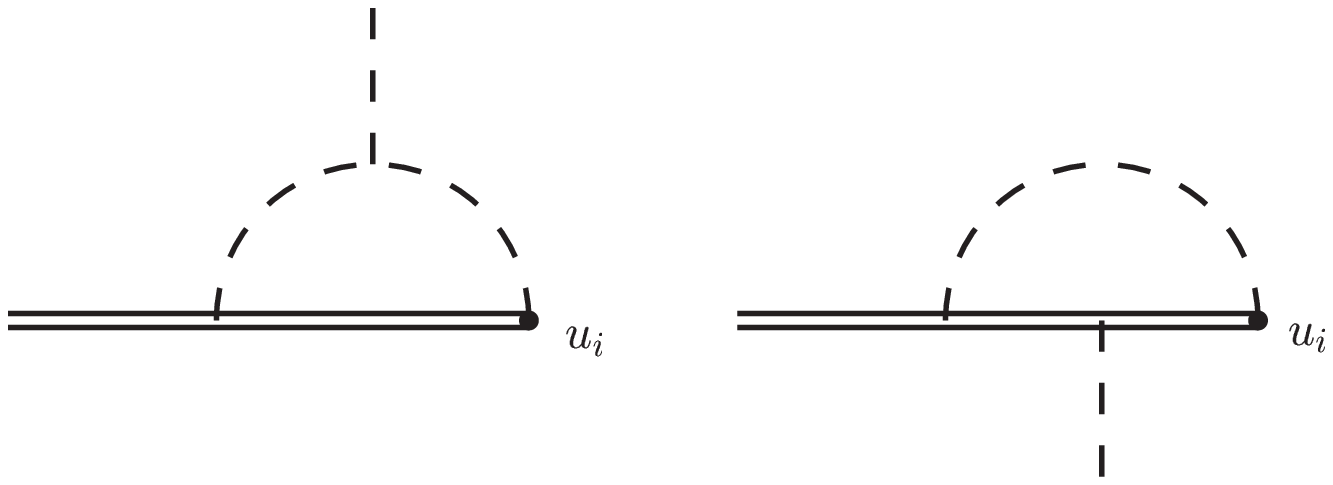}%
\caption{Graph with $ {\bf u}_i $ source, $ {\bf A}_i $ and $ {\bf c} $ lines.}
\end{center}
\end{figure}

\begin{figure}
\begin{center}
\includegraphics[width=8cm]{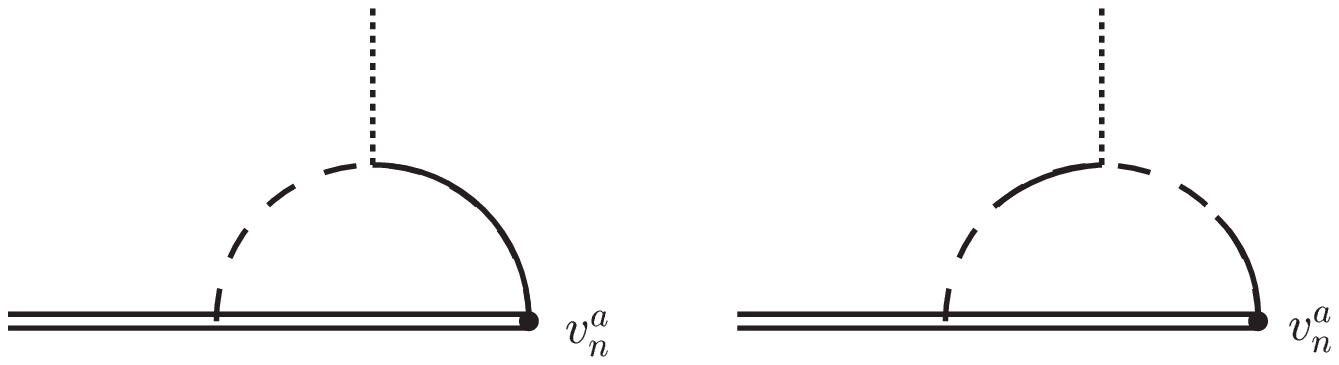}%
\caption{Diagram with $ {\bf v}_n $ source, $ {\bf A}_0 $ and $ {\bf c} $ lines.}
\end{center}
\end{figure}

\begin{figure}
\begin{center}
\includegraphics[width=5.5cm]{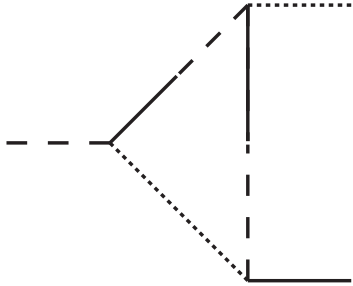}%
\caption{Graph contributing to the $ ({\bf A}_i{\bf E}_j{\bf A}_0) $ vertex function.}
\end{center}
\end{figure}

\begin{figure}
\begin{center}
\includegraphics[width=5.5cm]{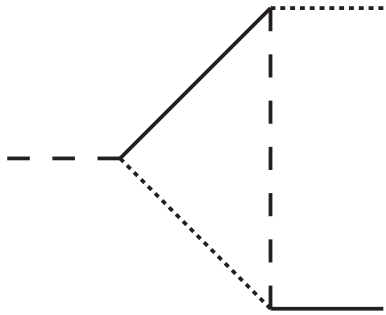}%
\caption{Graph with external gluon, Coulomb and $ {\bf E} $-field.}
\end{center}
\end{figure}

\begin{figure}
\begin{center}
\includegraphics[width=5.5cm]{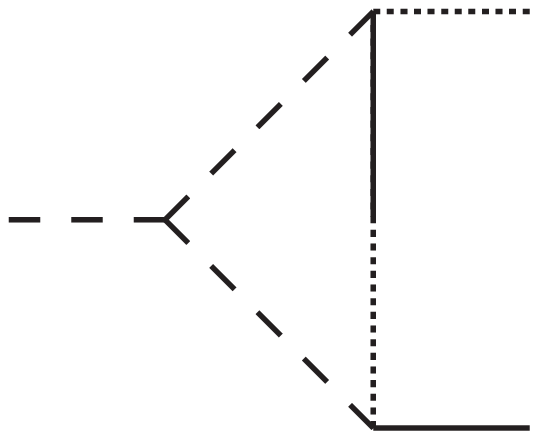}%
\caption{ Graph in the $ ({\bf A}_i{\bf E}_j{\bf A}_0) $ vertex function.}
\end{center}
\end{figure}

\end{document}